\documentclass[aps,prl,reprint,superscriptaddress,floatfix,10pt,showpacs]{revtex4-1}
\usepackage{graphicx}
\usepackage{dcolumn}
\usepackage{amsmath}
\usepackage{amssymb}
\pdfoutput=1
\usepackage{times}
\usepackage{epsfig}
\usepackage{color}
\usepackage{hyperref}

\begin{document}

\title{3D Modulated Spin Liquid model applied to URu$_2$Si$_2$}
\author{Christopher Thomas}
\affiliation{International Institute of Physics, Universidade Federal do Rio Grande do Norte, 59078-400 Natal-RN, Brazil}
\author{S\'{e}bastien Burdin}
\affiliation{Univ Bordeaux, LOMA, UMR 5798, F-33400 Talence, France, EU}         
\affiliation{CNRS, LOMA, UMR 5798, F-33400 Talence, France, EU}         
\author{Catherine P\'epin}
\affiliation{Institut de Physique Th\'eorique, CEA-Saclay, 91191 Gif-sur-Yvette, France}
\author{Alvaro Ferraz}
\affiliation{International Institute of Physics, Universidade Federal do Rio Grande do Norte, 59078-400 Natal-RN, Brazil}
\affiliation{Departamento de F\'isica Te\'orica e Experimental, Universidade Federal do Rio Grande do Norte,  59072-970 Natal-RN,Brazil}
\begin{abstract}

We have developed a 3D version for the Modulated Spin Liquid in a body-centered tetragonal lattice structure to describe the hidden order observed in URu$_2$Si$_2$ at $T_0\approx17.5$ K. This second order transition is well described by our model confirming our earlier hypothesis. The symmetry of the modulation is minimized for ${\bf Q}\equiv(1,1,1)$. We assume a linear variation of the interaction parameters with the lattice spacing and our results show good agreement with uniaxial and pressure experiments.
\end{abstract}

\date{\today}


\maketitle


The fascinating hidden order phase observed in the heavy fermion URu$_2$Si$_2$ below $T_0=17.5$ K~\cite{Palstra1985,Mydosh2011} is in the center of great discussion concerning the origin of the mechanism for this second order transition. From thermodynamics properties~\cite{Maple1986} a huge entropy quench is observed and this cannot be explained by a conventional antiferromagnetic transition due to the small value of the magnetic moment in this phase~\cite{Broholm1987}. 

The phase diagram obtained for the URu$_{2}$Si$_{2}$ is very interesting~\cite{Amitsuka2007,Hassinger2008,Villaume2008,Bourdarot2011}. In ambient pressure, the system undergoes a second order transition at $T_0=17.5$ K to a phase known as hidden order (HO), with a small magnetic moment $\mu\approx0.02$ $\mu_B$. At very small temperature, a superconducting phase is found below $T\approx1.5$ K. Applying hydrostatic pressure or uniaxial stress, the system  turns into an antiferromagnetic (AF) phase with a magnetic moment $\mu\approx0.4$ $\mu_B$ for $P_x\approx0.5$ GPa or $\sigma_x^a\approx0.33$ GPa, respectively. Bakker {\it et al.},~\cite{Bakker1992} showed that $T_0$ increases linearly with the uniaxial stress applied along the $a$ axis, and it decreases when the uniaxial stress is applied along the $c$ axis. Elastic neutron scattering measurements with uniaxial stress applied along [1,0,0], [1,1,0] and [0,0,1] directions show a very sensitive variation of the ordered moment~\cite{Yokoyama2005}. In-plane stress increases $\mu$, while a perpendicular stress does not change the ordered moment at all. 
The tuning parameter to the HO-AF transition seems to be the in-plane lattice constant $a$ as shown by Bourdarot {\it et al.}~\cite{Bourdarot2011}.
Inelastic neutron scattering (INS) measurements  show the formation of two gaps structure in URu$_2$Si$_2$: one at the incommensurate wave-vector $(1.4,0,0)$ and another at the commensurate $(1,0,0)$. The second gap is a good candidate for a signature of the HO, since the commensurate peak disappears in the AF phase~\cite{Broholm1987,Bourdarot2003,Wiebe2007,Villaume2008}. 
 

The various theories that have been presented along so far to explain the HO paradigm can be separated into sets of itinerant and localized models. Among the localized models we can cite the multipolar models~\cite{Cricchio2009,Haule2010,Harima2010,Toth2011,Kusunose2011}, and among the itinerant models there are different conjecture, indicating that the HO-AF transition is either driven by spin density wave~\cite{Ikeda1998,Mineev2005,Elgazzar2009}, by hybridization~\cite{Balatsky2009,Riseborough2012} or by orbital AF~\cite{Chandra2002}. 
The great advantage of our model is the ability to integrate in a natural way the HO and AF already in a realistic localized treatment, and to leave open the possibility to include charge fluctuation effects later on.

\begin{figure}[ht]
\centering
\includegraphics[width=0.9\columnwidth]{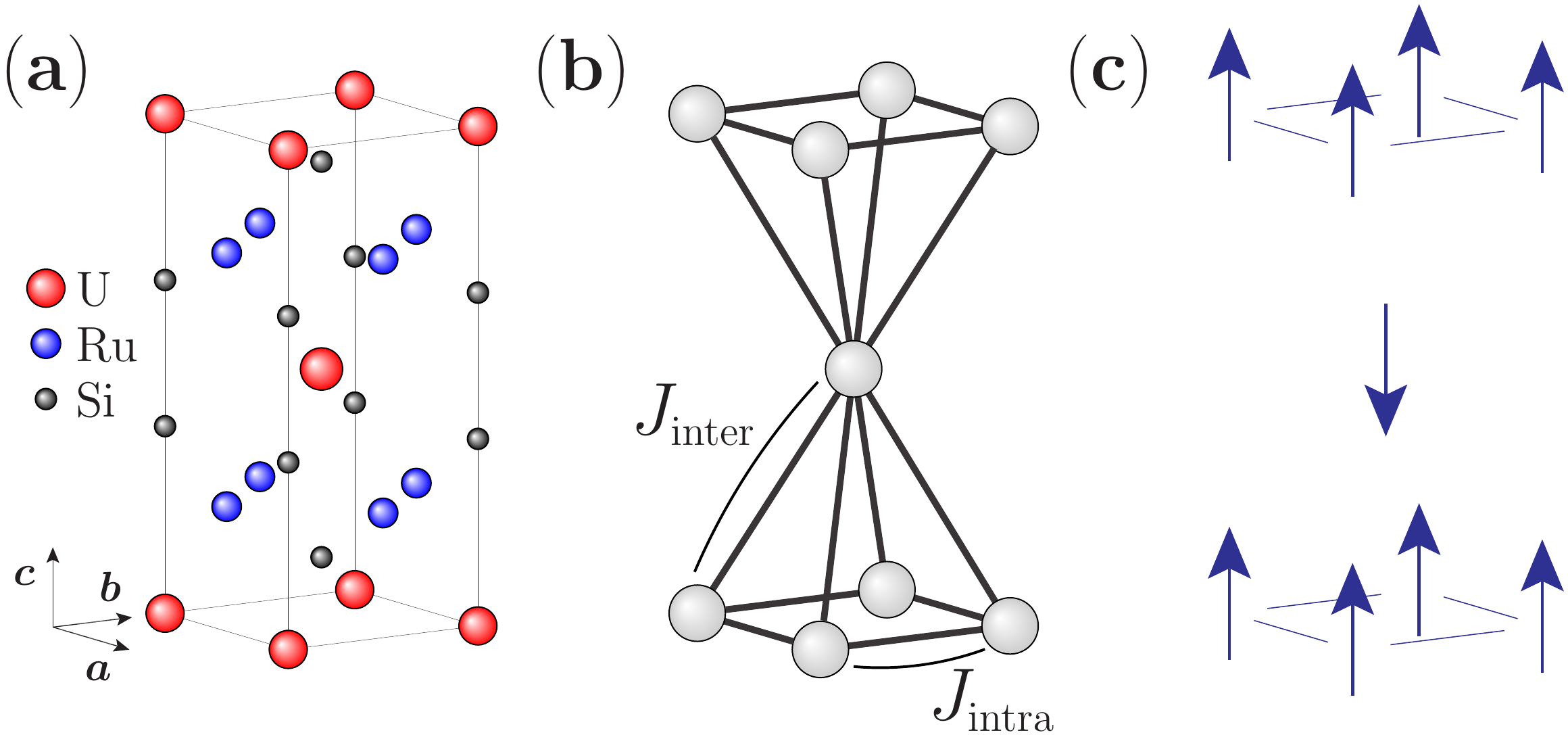}
\caption{\label{fig:together_3d}({\bf a}) The crystal structure of URu$_{2}$Si$_{2}$. The magnetism emerges from the $5f$ electrons of U atoms.
({\bf b}) The BCT-lattice Bravais structure of the uranium atoms. The intra-layer magnetic coupling is ferromagnetic, $J_{\text{intra}}<0$. The inter-layer coupling is antiferromagnetic, $J_{\text{inter}}>0$. 
({\bf c}) The structure of the magnetic long range order characterizing URu$_2$Si$_2$. 
}
\end{figure}

The Modulated Spin Liquid (MSL) in a 2D square lattice was developed in our previous work~\cite{Pepin2011} in order to explain the hidden order phase in URu$_2$Si$_2$. The 2D version of MSL model provides a simple scenario for the URu$_2$Si$_2$, with the HO phase resulting from a quantum phase transition where the local magnetic moments of the AF phase melt, restoring the time reversal symmetry, and preserving the lattice breaking symmetry. One general concept introduced by the MSL model is that the modulation of the SL reflects the structure of the magnetically ordered phase as shown in  FIG.~\ref{fig:together_3d}({\bf c}). The transport measurements indicate a continuous transition from HO to AF phase. The MSL is in this way the Resonant Valence Bond relative of the AF phase.

In the present work, we develop an extension for the MSL in a realistic 3D body-centered-tetragonal lattice (BCT-lattice) model. Our spin liquid (SL) framework in this 3D system is shown to be both realistic and experimentally motivated to explain the onset of the HO phase in URu$_2$Si$_2$.
The BCT-lattice structure of URu$_2$Si$_2$ is depicted in FIG. \ref{fig:together_3d}.
For convenience, we use here the tetragonal basis $({\bf a}, {\bf b}, {\bf c})$ which contains two U-atoms per unit cell of the tetragonal-lattice (T-lattice). 
Experimentally, the lattice parameters are $|{\bf a}|=|{\bf b}|=4.124$ \AA\ and $|{\bf c}|=9.5817$ \AA\ in the HO phase, with less than $1\%$ of variation when the system is warmed up to room temperature~\cite{Palstra1985}.

We start with a Heisenberg model Hamiltonian using the standard fermionic representation of quantum spins $1/2$, 

\begin{align}\label{Hamiltonian}
H_0=\sum_{\langle {\bf R},{\bf R'}\rangle,\sigma\sigma'}
J_{\bf RR'}\chi_{{\bf R}\sigma}^{\dagger}\chi_{{\bf R}\sigma'}\chi_{{\bf R'}\sigma'}^{\dagger}\chi_{{\bf R'}\sigma}~,   
\end{align} 
where the fermion annihilation (creation) operators $\chi_{{\bf R}\sigma}^{(\dagger )}$ satisfy the local constraints $\sum_{\sigma=\uparrow , \downarrow}\chi_{{\bf R}\sigma}^{\dagger}\chi_{{\bf R}\sigma}=1$. For simplicity, we consider only the magnetic interaction between nearest neighbor sites ${\bf R}$ and ${\bf R'}$, as depicted in  FIG.~\ref{fig:together_3d}({\bf b}). We assume this nearest neighbor interaction to be ferromagnetic ($J_{\bf RR'}=J_{\text{intra}}<0$ ) inside each layer and antiferromagnetic ($J_{\bf RR'}=J_{\text{inter}}>0$) in between adjacent layers. This is the simplest and the most natural interaction which can reproduce the magnetically ordered phase obtained experimentally at high pressure~\cite{Amitsuka2007}: an intra-layer ferromagnetism together with inter-layer antiferromagnetism (see FIG.~\ref{fig:together_3d}({\bf c})). 

Generalizing the procedure of Ref.~\onlinecite{Pepin2011}, the Heisenberg Hamiltonian~(\ref{Hamiltonian}) is decoupled for each bond ${\bf RR'}$ using appropriated Hubbard-Stratonovich transformations. We find the following Lagrangian
 
\begin{align}
{\cal L}_{0} &= \sum_{{\bf R}\sigma}\chi_{{\bf R}\sigma}^{\dagger}
\left( \partial_{\tau}+\lambda_{\bf R}+\sigma\sum_{\bf z}m_{\bf R+z}\right)\chi_{{\bf R}\sigma}\notag\\
&\hspace{-0.1cm}-\sum_{\bf R}\lambda_{\bf R}+
\sum_{n\sigma}\sum_{\langle {\bf R}\in{ L}_n, {\bf R'}\in{ L}_{n\pm 1}\rangle }[\varphi_{\bf RR'}\chi_{{\bf R}\sigma}^{\dagger}\chi_{{\bf R'}\sigma}
+c.c.]\notag\\
&\hspace{-0.5cm}+\sum_{n}\left(\sum_{\langle {\bf R}\in{ L}_n, {\bf R'}\in{ L}_{n\pm 1}\rangle }\frac{2|\varphi_{\bf RR'}|^2}{J_{\text{inter}}}-\sum_{\langle {\bf R},{\bf R'}\rangle\in{ L}_n}\frac{m_{\bf R}m_{\bf R'}}{2J_{\text{intra}}}\right),
\label{Lagrangian}
\end{align}
where, ${L}_n$ denotes the layer $n$, and the sum over ${\bf z}$ refers to the nearest neighbors within the same layer. In the following, the Hubbard-Stratonovich fields will be replaced by their constant, self-consistent, mean-field expressions, $\varphi_{\bf RR'}=-J_{\text{inter}}\sum_{\sigma}\langle\chi_{{\bf R}\sigma}^{\dagger}\chi_{{\bf R'}\sigma}\rangle$ and $m_{\bf R}=J_{\text{intra}}\sum_{\sigma}\sigma\langle \chi_{{\bf R}\sigma}^{\dagger}\chi_{{\bf R}\sigma}\rangle$. 

Note that this magnetic, intra-layer only, decoupling channel leads to a degenerate mean-field system, with each layer becoming effectively ferromagnetic, but with an easy axis completely decoupled from the other layers. This degeneracy does not distinguish an artificially fully ferromagnetic order from the expected AF order depicted by  FIG.~\ref{fig:together_3d} ({\bf c}). 
A more general decoupling scheme would consist in splitting arbitrarily the  inter-layer interaction, $J_{\text{inter}}\equiv J_{\text{SL}}+J_{\text{AF}}$, following closely the procedure used in  Ref.~\onlinecite{Pepin2011}: the terms  with $J_{\text{AF}}$ and $J_{\text{SL}}$ are decoupled in the magnetic and SL channels respectively. At mean-field level, the degenerescence is lifted by the contribution from the inter-layer part of the local Weiss fields, i.e., the contribution originating from $J_{\text{AF}}$ terms. Despite an apparent higher complexity, this generalized mean-field problem is formally identical to the one described originally by the Lagrangian~(\ref{Lagrangian}). Indeed, considering the BCT-lattice coordination numbers, this general decoupling scheme can be derived at mean-field level from the one used here by simply mapping $J_{\text{intra}}\mapsto J_{\text{intra}}+2J_{\text{AF}}$, and $J_{\text{inter}}\mapsto  J_{\text{SL}}=J_{\text{inter}}-J_{\text{AF}}$. 
As we will see next, what is remarkable here, with the BCT-lattice, is that the competition between the magnetic and the MSL orders is simply tunable by changing the ratio $( J_{\text{intra}}+2J_{\text{AF}})/(J_{\text{inter}}-J_{\text{AF}})$, which is qualitatively independent from the arbitrary splitting if we take $J_{\text{inter}}= J_{\text{SL}}+J_{\text{AF}}$. Therefore, in this work, we just assume $J_{\text{AF}}=0$ and we consider $J_{\text{intra}}/J_{\text{inter}}$ as the new tuning parameter which is phenomenologically associated to pressure variations. 

Experimentally, pressure has a direct effect on the ratio between the inter-layer magnetic coupling, $J_{\text{inter}}$, and the intra-layer one, $J_{\text{intra}}$. This is not standard for heavy-fermion systems, where pressure variations may often change the local energy level of the $f$ electrons. This different phenomenological approach is supported here by a strong experimental evidence: in URu$_2$Si$_2$, pressure favors a magnetic phase. Here, of course, the mechanism is not Doniach-like.

We introduce the Fourier transform of the fields, 
$\chi_{{\bf k}\sigma}\equiv\frac{1}{\sqrt{N}}\sum_{\bf R}e^{-i{\bf k}.{\bf R}}\chi_{{\bf R}\sigma}$, $m_{\bf k}\equiv\frac{1}{\sqrt{N}}\sum_{\bf R}e^{-i{\bf k}.{\bf R}}m_{\bf R}$, $\varphi_{\bf q}\equiv\frac{e^{i\theta_{\bf q}}}{2\sqrt{N}}\sum_{n}\sum_{\langle{\bf R}\in L_{n}{\bf R'}\in L_{n+1}\rangle }e^{-i{\bf q}.\left(\frac{{\bf R}+{\bf R'}}{2}\right)}\varphi_{{\bf R}{\bf R'}}$, where $N$ is the number of lattice sites. The phase factor $\theta_{\bf q}\equiv{\bf q}\cdot{\bf R}_0$ is introduced in order to fix the origin of the bond lattice at real space position ${\bf R}_0\equiv ({\bf a}+{\bf b}+{\bf c})/4$.

\begin{figure}[ht]
\centering
\includegraphics[width=0.8\columnwidth]{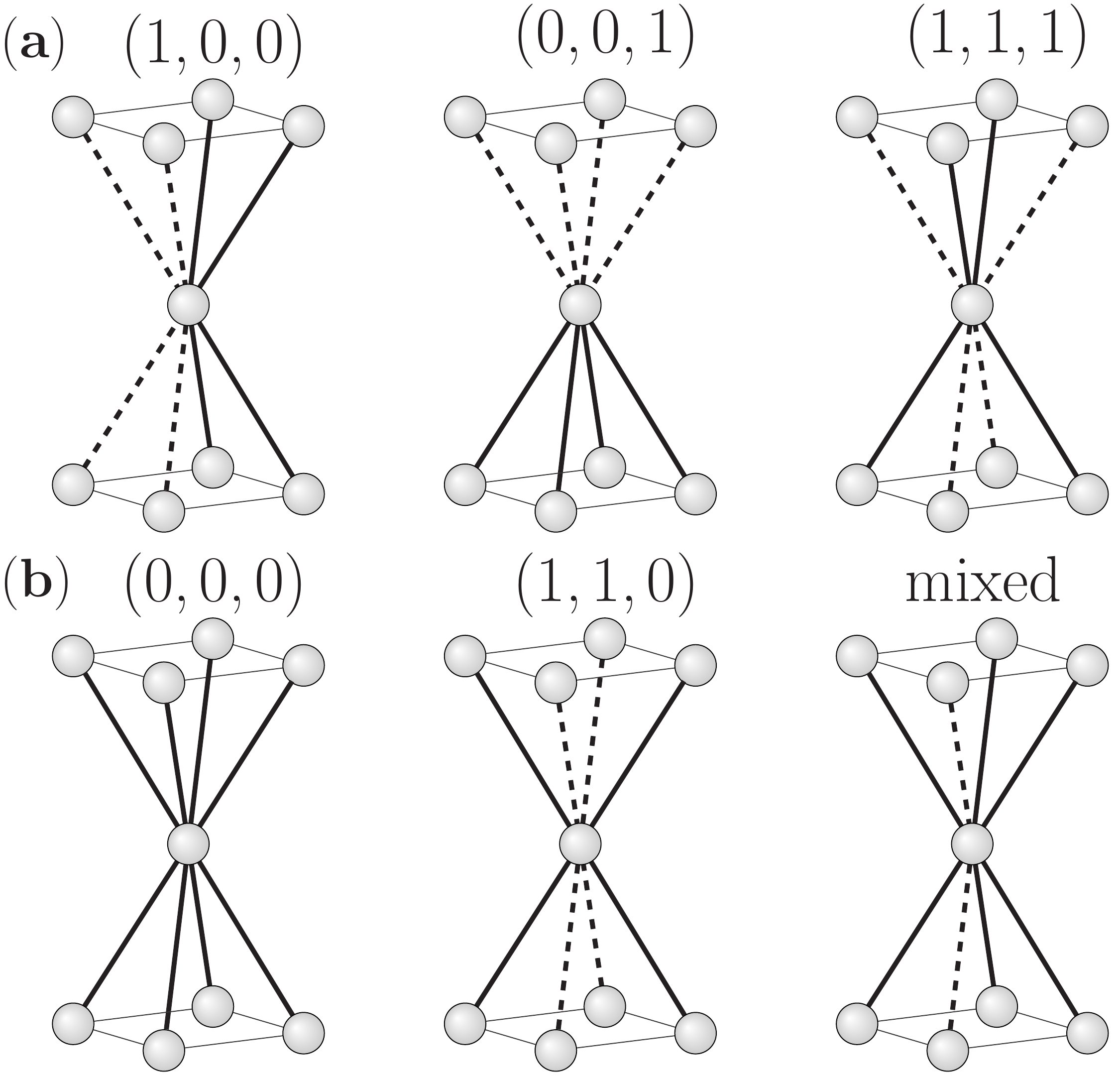}
\caption{Possible SL long range ordered modulations represented on the unit cell of the T-lattice. When it is well defined, the associated wave-vector is indicated in reduced coordinates, $(h,k,l)$. Dotted lines represent $\varphi_{\bf RR'}=\varphi_0+\varphi_{\bf Q}$, and solid lines represent $\varphi_{\bf RR'}=\varphi_0-\varphi_{\bf Q}$ bond modulation. ({\bf a}) Here we show the three different relevant odd MSL wave-vectors considered. ({\bf b}) Respectively the homogeneous SL, an even wave-vector that we exclude (see text), and a mixed modulation which may invoke more than one wave-vector (not considered in this work). } 
\label{Fig_URS_3Dmodulations}
\end{figure}

Hereafter we will concentrate our analysis on the following mean-field parameters: the uniform SL, $\Phi_{0}\equiv \varphi_{(0,0,0)}/\sqrt{N}$, the modulated SL, $\Phi_{\bf Q}\equiv \varphi_{\bf Q}/\sqrt{N}$, and the N\'eel staggered magnetization AF, $S_{{\bf Q}_{\text{AF}}}\equiv m_{{\bf Q}_{\text{AF}}}/\sqrt{N}$. There is an important difference with the square lattice MSL~\cite{Pepin2011}. Here, the equivalences between wave-vectors ${\bf Q}$ for the $\varphi$ fields do not refer to the same Brillouin zones as the ones between ${\bf Q}_{\text{AF}}$ for the $m$ fields. This is due to the fact that the magnetization fields are defined on the sites, while the SL fields are defined on the bonds. 
The symmetry group of the AF phase corresponds to a T-lattice, and ${\bf Q}_{\text{AF}}$ is thus defined {\it modulo} the first Brillouin zone of the T-lattice. 
We consider that the MSL states, similarly to what happens in the AF phase, satisfy the T-lattice translational symmetries, although different ${\bf Q}$ modulation vectors are defined {\it modulo} a larger Brillouin zone, characterizing the long range order with T-lattice periodicity, but with different intra-shell symmetry breaking (see FIG.~\ref{Fig_URS_3Dmodulations}).

For simplicity, we consider at most the case of one kind of modulation at a time in the presence of an homogeneous solution. If ${\bf Q}$ is the wave vector associated with the MSL, ${\bf Q}_{\text{AF}}$  is the wave vector associated with the AF. We get: $m_{\bf R}=S_{{\bf Q}_{\text AF}} e^{i {\bf Q}_{\text{AF}} \cdot {\bf R}}$, $\varphi_{\bf q}=\Phi_{0}\sqrt{N}\delta_{\bf q }+\Phi_{\bf Q}\sqrt{N}\delta_{{\bf q}+{\bf Q}}$, where $\delta$ denotes the Kronecker delta. Note that a completely equivalent Ansatz can also be made in the direct bond lattice, similarly to what was done earlier in~\cite{Pepin2011}, namely,  $\varphi_{\bf RR'}=\delta_{{\bf R}, {\bf R'}+{\bf z}}\left(\varphi_0+\varphi_{\bf Q}e^{-i {\bf Q}\cdot({\bf R}+{\bf R'})/2 - i\theta_{\bf Q}}\right)$. 

We introduce the following reduced notation for the SL modulation wave-vectors: ${\bf Q}\equiv (h,k,l)$. The parity of the modulation can be obtained from the phase factor $e^{i\pi(h,k,l)}=\pm 1$. The sign $+$ ($-$) characterizes wave-vectors with even (odd) parity. The only possible MSL characterized by a single modulation wave-vector have odd parity. The definition of parity can be extended to the AF wave-vector. We find here that the parity of ${\bf Q}_{\text{AF}}$ is odd (see FIG.~\ref{fig:together_3d}({\bf c})).
For simplification, we consider here, MSL wave-vectors with either $1$ or $3$ modulations. Due to the ${\bf a}\rightarrow {\bf b}$ symmetry, we finally need to compare only three types of modulations (see FIG.~\ref{Fig_URS_3Dmodulations}). All these wave-vectors break BCT-lattice symmetry and have the 
periodicity of a T-lattice: ${\bf Q}_1\equiv(1,0,0)$ also breaks a rotation symmetry and characterizes an orthorhombic lattice, ${\bf Q}_2\equiv(0,0,1)$ may break a mirror symmetry, and ${\bf Q}_3\equiv (1,1,1)$ clearly belongs to the T-lattice group.

The selection between different modulation vectors ${\bf Q}$ is obtained by comparing the corresponding minimized free-energies per site, which is given by
\begin{align}\label{eq:fren}
F&(\lambda_{0},\Phi_{0},\Phi_{\bf Q},S_{{\bf Q}_{\text{AF}}})=-\frac{k_{B}T}{N}\sum_{\bf k}\sum_{\alpha=\pm}\ln{\left[ 1+e^{-\beta\Omega_{\bf k}^{\alpha}}\right] } 
\nonumber\\
&-\lambda_{\bf 0}+\frac{4}{J_{\text{inter}}}\left[|\Phi_{0}|^2+|\Phi_{\bf Q}|^2\right]
-\frac{2}{J_{\text{intra}}}|S_{{\bf Q}_{\text{AF}}}|^2~,
\end{align}
where the sum over ${\bf k}$ is taken over the full Brillouin zone and the eigenergies are given by 
\begin{align}
\Omega_{\bf k}^{\pm}=
\lambda_{0}
\pm 4\sqrt{S_{{\bf Q}_{AF}}^{2}+4\Phi_{0}^{2}\gamma_{1,{\bf k}}^2
+4|\Phi_{\bf Q}|^{2}\gamma_{2,{\bf k},{\bf Q}}^2
}\,. 
\end{align}
with $\gamma_{1,{\bf k}} = \cos{\left(\frac{k_ a}{2}\right)}\cos{\left(\frac{k_b}{2}\right)}\cos{\left(\frac{k_c}{2}\right)}$ and $\gamma_{2,{\bf k},{\bf Q}} = \cos{\left(\frac{k_ a}{2}+\frac{Q_a}{4}\right)}\cos{\left(\frac{k_b}{2}+\frac{Q_b}{4}\right)}\cos{\left(\frac{k_c}{2}+\frac{Q_c}{4}\right)}$.
For a given modulating wave-vector ${\bf Q}$, the staggered magnetization, $S_{{\bf Q}_{\text{AF}}}$, and the homogeneous and modulated SL parameters, $\Phi_{0}$ and $\Phi_{\bf Q}$ are obtained directly from the minimization of the free-energy function. The free energy is calculated minimizing eq. \eqref{eq:fren}, using Powell's method~\cite{Press2007}
, with the auxiliary equation to fix the number of $n_f$, $1=\frac{1}{N}\sum_{\bf k}\sum_{\alpha=\pm}\frac{1}{ 1+e^{\beta\Omega_{\bf k}^{\alpha}}}$.

\begin{figure}[ht]
\centering
\includegraphics[width=1\columnwidth]{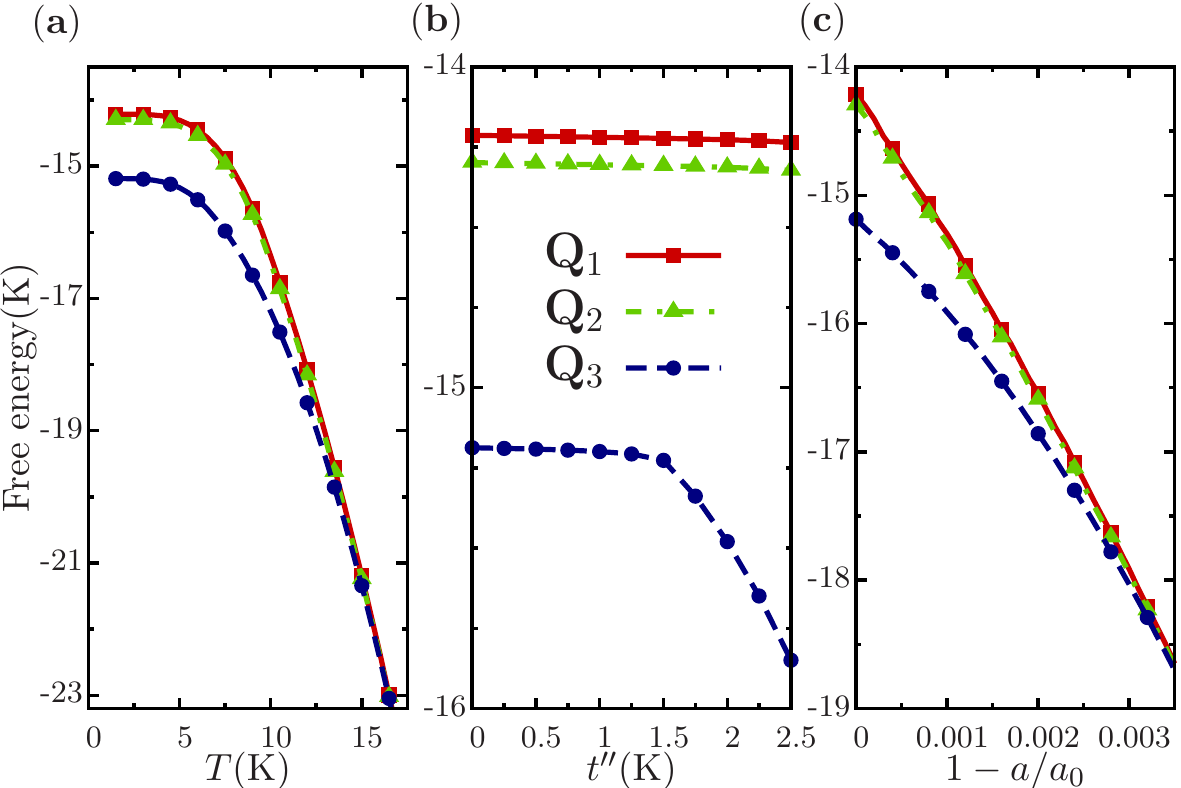}
\caption{\label{fig:min_free_varia} Free energy of three modulating vectors ${\bf Q}_1=(1,0,0)$, ${\bf Q}_2=(0,0,1)$ and ${\bf Q}_3=(1,1,1)$ as a function of: ({\bf a}) $T$ for $t''=0$ K and $a=a_0$; ({\bf b}) {\it hopping} for $T=2$ K and $a=a_0$; and ({\bf c}) $1-a/a_0$  for $T=2$ K and $t''=0$ K. The parameters used were $J_{\text{intra}}(a_0)=-6.5$ K, $J_{\text{inter}}(a_0)=37$ K and $\mathcal{B}=800$ K, as described in text.}
\end{figure}

FIG.~\ref{fig:min_free_varia} depicts the behavior of the free energy for the three different wave vectors ${\bf Q}_1$, ${\bf Q}_2$ and ${\bf Q}_3$. We find, by varying the different physical parameters of the model, that the minimum is always obtained for the wave vector ${\bf Q}_3=(1,1,1)$.
These states, which correspond to the space group No. 134 $P4_2/nnm$ also appear to be compatible with the crystallographic analysis of URu$_2$Si$_2$ made by Harima {\it et al.}~\cite{Harima2010}.

A n.n.n. {\it hopping} $t''$ may be included  in order to phenomenologically take into account some frustration and intra-plane spin-liquid contribution.
In this case, the intra-layer term is the same for momenta ${\bf k}$ and ${\bf k+Q}$, and the extension of our model is directly obtained if we perform the change  $\Omega_{\bf k}^{\pm}\mapsto \Omega_{\bf k}^{\pm}+t''\Phi_0\cos{\left(k_a\right)}\cos{\left(k_b\right)}$. 

Motivated by the pressure experiments which are specially dedicated to the anisotropy and the uniaxial effects~\cite{Bakker1992,Yokoyama2005,Bourdarot2011}, we relate our microscopic interaction with pressure.
Bourdarot {\it et. al.}~\cite{Bourdarot2011} showed that the uniaxial stress is the relevant variational parameter to change the behavior of the system. They considered that the deformation is in the linear elastic regime.
In this work, we propose that our parameters $J_{\text{intra}}$ and $J_{\text{inter}}$ also vary linearly with the lattice parameter $a$. The variation can be simply  written as $J_{\text{inter}}(a)=J_{\text{inter}}(a_0)+\mathcal{B}_1(1-a/a_0)$ and $J_{\text{intra}}(a)=J_{\text{intra}}(a_0)-\mathcal{B}_2(1-a/a_0)$, where $a_0$ is the value of lattice parameter in ambient pressure.

Here, the fitting parameters chosen to produce a phase diagram in qualitatively good agreement with experiment are: $J_{\text{inter}}(a_0)=37$ K is chosen to obtain $T_0=17.5$ K and $J_{\text{intra}}(a_0)=-6.5$ K is chosen in order to obtain the best approximated value for the critical stress $1-a/a_0\approx1.45\times10^{-3}$~\cite{Bourdarot2011}. To have a good agreement with experiment, we also choose the linear coefficient of $J_{\text{inter}}(a)$, $\mathcal{B}_1$, to have the same slope of $T_0$, as observed experimentally, and we define $\mathcal{B}_2=\mathcal{B}_1\equiv\mathcal{B}$ for simplicity. Both $J_{\text{inter}}(a)$ and $J_{\text{intra}}(a)$ increase their absolute values when $1-a/a_0$ increases. 
Our choice of the interaction parameters variation is, of course, a simplified view of the experiment: if we apply an uniaxial stress, the in-plane lattice parameters become different. In our case both in-plane parameters decrease in the same way.    

\begin{figure}[ht]
\includegraphics[width=0.8\columnwidth]{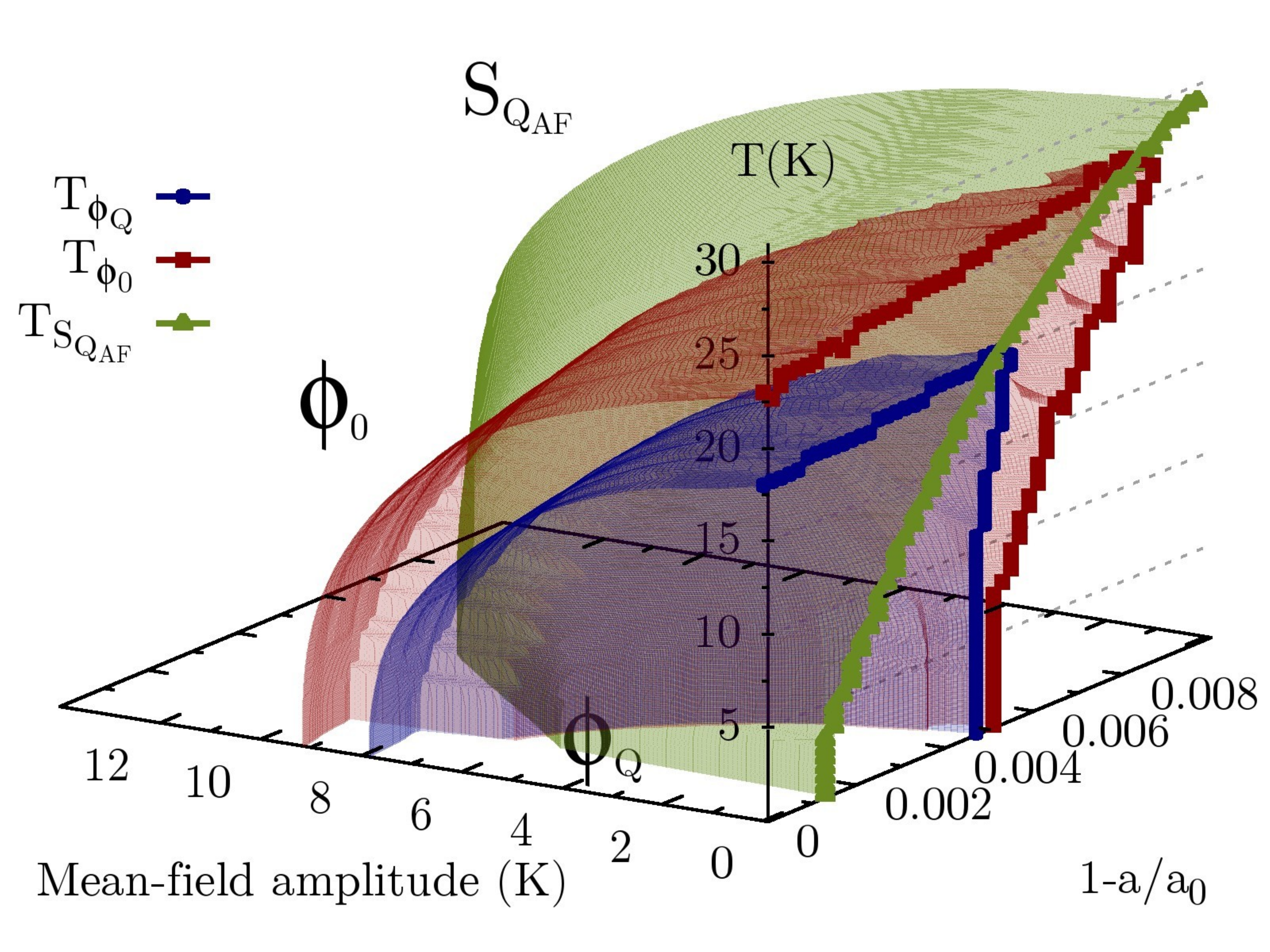}
\caption{\label{fig:diagram_3D} The phase diagram for the modulation ${\bf Q}_3$ as function of deformation ($1-a/a_0$) and $T$. Inside the surfaces the mean-field amplitudes are different from zero. There is a region where the three order parameters coexist. The circle (red), square (blue) and triangle (green) lines represent the critical temperatures for the parameters $\Phi_{\bf Q}$, $\Phi_{0}$ and $S_{{\bf Q}_{\text{AF}}}$, respectively. The parameters used are: $t''=-2.5$ K, $J_{\text{intra}}(a_0)=-6.5$ K, $J_{\text{inter}}(a_0)=37$ K and $\mathcal{B}=800$ K. }
\end{figure}

The resulting phase diagram 
is shown in FIG. \ref{fig:diagram_3D}. The variation on $1-a/a_0$ shows very good agreement with the experimental results.
We define $T_{\Phi_{\bf Q}}$, $T_{\Phi_0}$ and $T_{S_{\bf Q_{\text{AF}}}}$ as the critical temperatures for the parameters $\Phi_{\bf Q}$, $\Phi_{0}$ and $S_{{\bf Q}_{\text{AF}}}$, respectively.
Increasing $1-a/a_0$, the MSL critical temperature $T_{\Phi_{\bf Q}}$, increases linearly until it reaches the AF ordering temperature $T_{S_{\bf Q_{\text{AF}}}}$ and then it goes to zero showing a re-entrance behavior due the presence of the hopping $t''$. The homogeneous component $T_{\Phi_{0}}$, shows a similar variation, although with a bigger amplitude for $t''$ different from zero. In our model, $\Phi_0$ persists for big values of $T$ or $1-a/a_0$, but with a small intensity. For the sake of simplicity, we define $T_{\Phi_0}$ when $\Phi_0=0.6$ K. The $T_{S_{\bf Q_{\text{AF}}}}$ also increases with $1-a/a_0$, but it shows two different behaviors: at first, a fast increase when it is inside the SL phase 
and, a linear increase outside the SL. Here the effect of hopping is visible: without this effect, $T_{S_{\bf Q_{\text{AF}}}}$ will present just linear slopes. A first order transition can also be obtained for $T_{S_{\bf Q_{\text{AF}}}}$ inside the MSL phase when $t''$ is present.

In conclusion, we have developed a modulated spin liquid model in the realistic three-dimensional BCT-lattice. This provides a simple scenario for URu$_2$Si$_2$, where the hidden order results from a quantum phase transition with a very unusual behavior: the magnetic moments of the AF phase melt at 
 low pressure, restoring the time reversal symmetry, but the lattice symmetry breaking is still present. 
We analyzed how this SL melting in a BCT-lattice can lead to different modulation wave-vectors, among which ${\bf Q}_3=(1,1,1)$ is found to be the most stable energetically.
The theoretical phase diagram reproduces qualitatively well what is observed experimentally for URu$_2$Si$_2$. We identify a second order transition at $T_0\approx17.5$ K and a first order transition from the MSL phase to the AF phase at low temperature. The linear dependence of $J_{\text{intra}}$ and  $J_{\text{inter}}$ with the variation $1-a/a_0$ is a key point of our study, which is confirmed by experimental results~\cite{Bourdarot2011}. Our results clearly show that the choice of an appropriate modulation vector is crucial for the stability of the MSL phase. This could be directly checked experimentally by INS measurements. By comparing all crystallographic directions one could find clear evidence for what this preferable modulation might be. Raman scattering experiments could also provide another independent check of our results since the orientation dependence of Raman spectrum could establish if the modulation is indeed characterized or not by our ${\bf Q}_3$ vector. We believe that our study is a very good test for a MSL paradigm.

\begin{acknowledgments}
We acknowledge the financial support of Capes-Cofecub Ph 743-12.
This research was also supported in part by the Brazilian Ministry of Science, Technology and Innovation (MCTI) and the Conselho Nacional de Desenvolvimento Cientifico e Tecnol\'ogico (CNPq).
Finally, we thank M. R. Norman for his help and for innumerous discussions and F. Bourdarot for  carefully reading an earlier version of this manuscript and for very usefull comments. We also acknowledge C. Lacroix, P. Coleman, H. Harima and K. Miyake for discussions.
\end{acknowledgments}
 
\bibliography{bibliografia02}

\end{document}